# Pressure-induced phase transition for single crystalline $LaO_{0.5}F_{0.5}BiSe_2$

Masaya Fujioka[1,a)], Masashi Tanaka[1], Saleem J. Denholme[1], Takuma Yamaki[1,2], Hiroyuki Takeya[1], Takahide Yamaguchi[1], and Yoshihiko Takano[1,2]

[1] *National Institute for Materials Science, 1-2-1 Sengen, Tsukuba, Ibaraki 305-0047, Japan*

[2] *University of Tsukuba, 1-1-1Tennodai, Tsukuba, Ibaraki 305-0001, Japan*

We have demonstrated a pressure-induced phase transition from a low-$T_c$ phase to a high-$T_c$ phase in a single crystal of the superconductor $LaO_{0.5}F_{0.5}BiSe_2$. The high-$T_c$ phase appears at 2.16 GPa and the maximum superconducting transition temperature ($T_c$) is observed at 6.7 K under 2.44 GPa. Although the anisotropy ($\gamma$) for the low-$T_c$ phase is estimated to be 20, it is reduced by around half (9.3) in the high-$T_c$ phase. This tendency is the same for the $BiS_2$ system. The $T_c$ of $LaO_{0.5}F_{0.5}BiSe_2$ has continued to increase up to the maximum pressure of this study (2.44 GPa). Therefore applied further pressure has the potential to induce a much higher $T_c$ in this system.

___________________________

[a)] Author to whom correspondence should be addressed. Electronic mail: FUJIOKA.Masaya@nims.go.jp

Recently, a series of BiS$_2$-based superconductor such as Bi$_4$O$_4$S$_3$[1], LnO$_{1-x}$F$_x$BiS$_2$ (Ln = La, Ce, Pr, Nd, Yb) [2-8] and Sr$_{1-x}$F$_x$BiS$_2$ [9,10] were discovered. These compounds have a layered structure composed of superconducting BiS$_2$ layer and charge reservoir blocking layers. This structure is in common with the cuprate superconductors [11] and Fe-based superconductors [12]. The most attractive point of LnO$_{1-x}$F$_x$BiS$_2$ is the pressure induced phase transition from a low-$T_c$ phase to a high-$T_c$ phase[2, 13-16]. For instance, the $T_c$ of LaO$_{0.5}$F$_{0.5}$BiS$_2$, which was synthesized by a conventional solid state reaction, is obtained at only 3 K. On the other hand, the $T_c$ largely increases above 10 K by using high pressure synthesis. Furthermore, even if the sample is prepared by using a conventional solid state reaction method, high pressure measurements can enhance the $T_c$ above 10 K [16]. Very recently, we have reported the pressure induced phase transition for the single crystalline PrO$_{0.5}$F$_{0.5}$BiS$_2$ [17]. From this study, it was found that the semiconducting-like behavior in low-$T_c$ phase changes to the metallic behavior and the $\gamma$ is reduced by around half after the phase transition.

The modification of superconducting and blocking layers is one of the promising approaches to discover new superconducting materials [18]. Polycrystalline LaO$_{0.5}$F$_{0.5}$BiSe$_2$ was discovered by the replacement of the BiS$_2$ superconducting layers in LaO$_{1-x}$F$_x$BiS$_2$ to the BiSe$_2$ layers [19]. In such a material whether the pressure induced phase transition can be observed or not is a point of interest. Also it will provide useful information to know the mechanism of emergence for the high-$T_c$ phase. In previous research, we have succeeded in preparing single crystalline LaO$_{0.5}$F$_{0.5}$BiSe$_2$ and performing structural analysis of this compound [20]. According to our previous study on PrO$_{0.5}$F$_{0.5}$BiS$_2$[17], it was found that applying uniaxial pressure to single crystals effectively induces the high-$T_c$ phase. Also, measurements using single crystals are suitable to investigate the intrinsic properties of these materials. In this research, we investigated the pressure induced phase transition of single crystalline LaO$_{0.5}$F$_{0.5}$BiSe$_2$.

The single crystals were prepared by the CsCl flux method. The details of characterization are shown in ref. 20. All described fluorine content in this article is the nominal amount. X-ray diffraction (XRD) measurement showed that the single crystals were highly oriented to the $c$-axis as shown in Fig.1. The $c$ lattice parameter was estimated to be 1.414(4) nm. Electrical resistivity was measured by a physical properties measurement system (PPMS) under varying pressures up to a maximum of 2.44 GPa. The measurement was performed by a standard four-probe method from 300 K to 2 K in a piston-cylinder-type pressure cell. The measured sample was formed into rectangles with dimensions of around 1.3 × 0.4 × 0.02 mm². Since the thickness (20 μm) is smaller compared to the other dimensions, therefore the actual applied pressure should be almost uniaxial along the $c$-axis, despite the fact that we are using a hydrostatic pressure technique. This relationship between sample configuration and anisotropic pressure has been demonstrated previously in ref.21. The resistivity measurements were performed under several magnetic fields with directions both parallel to the $ab$ plane ($H//ab$) and $c$ axis ($H//c$), under 1.17 and 2.44 GPa.

The temperature dependence of resistivity for $LaO_{0.5}F_{0.5}BiSe_2$ under various pressures is shown in figure 2 (a) and (b). The data points at ambient pressure were noisy due to poor contact resistance, but was improved by applied pressure. In this study, $T_c^{onset}$ and $T_c^{zero}$ were regarded as described in figure 2 (b). The resistivity at ambient pressure and 0.76 GPa starts to decrease at 3.4 K and a second drop is observed at 2.4 K. Although the zero resistivity was not obtained at ambient pressure, $T_c^{zero}$ appeared at 2.2 K under 0.76 GPa. Additionally, a small hump at around 40 K is observed at ambient pressure and 0.76 GPa. This hump was also confirmed in our previous report[22]. When increasing pressure up to 1.17 GPa, it completely disappears. However, even though the pressure increases up to 2.16 GPa, $T_c^{zero}$ does not increase above 2.7 K. On the other hand, at 2.16 GPa, the resistivity starts to drop at 6.3 K and shows a step like behavior. This means that some of parts in the sample form the high-$T_c$ phase. For further pressure up to 2.44 GPa, the $T_c^{onset}$ clearly appears at 6.7 K and the $T_c^{zero}$ suddenly increases up to 5 K. From this figure, it is expected that the drop of resistivity at around 3 K corresponds to the low-$T_c$ phase, and the high-$T_c$ phase is obtained at around 7 K. However, under ambient pressure and 0.76 GPa, another drop of resistivity is also observed at 2 K. This may be due to the inhomogeneous fluorine concentration, or a local distortion for the extreme thinness of the sample.

In this measurement, we confirmed the discontinuous increase in $T_c$ of LaO$_{0.5}$F$_{0.5}$BiSe$_2$, as was also observed with the BiS$_2$ system. Therefore, regardless of whether the superconducting layers are made up of sulfur or selenium, a pressure induced phase transition occurs. Furthermore, in the high-$T_c$ phase, the BiS$_2$ and BiSe$_2$ systems show similar behavior. Namely, the resistivity in the normal conducting state of the high-$T_c$ phase gradually increases with increasing applied pressure[17]. On the other hand, in the low-$T_c$ phase, although the BiS$_2$ system shows semiconducting-like behavior, the metallic behavior is observed in the BiSe$_2$ system.

The temperature dependences of resistivity under magnetic field along the *c*-axis and the *ab* plane were measured as shown in figure 3 (a)-(d). When the magnetic field is parallel to the *c* axis, superconductivity is drastically suppressed with increasing magnetic field. This is typical behavior observed in superconductors with a layered structure. Figure 4 shows the upper critical field ($H_{c2}$) versus temperature for the low- and high-$T_c$ phase. To estimate the $H_{c2}$, 90 % of resistivity at $T_c^{onset}$ is adopted as a criterion. In this article, $H_{c2}$ obtained from magnetic field $H//ab$ and $H//c$ are described as $H_{c2}^{//ab}$ and $H_{c2}^{//c}$ respectively. In addition, the $\gamma$, the coherence length at 0 K along the *c* axis ($\xi_c(0)$) and the *ab*-plane ($\xi_{ab}(0)$) are estimated from figure 4 by using following formula: $\gamma = H_{c2}^{//ab} / H_{c2}^{//c}$, $H_{c2}^{//c} = \Phi_0/2\pi\xi_{ab}^2(0)$ and $H_{c2}^{//ab} = \Phi_0/2\pi\xi_{ab}(0)\xi_c(0)$, where $\Phi_0$ is flux quantum. Obtained these superconducting parameters in low- and high-$T_c$ phase are summarized in table I.

Although the $\gamma$ for the low-$T_c$ phase is estimated to be 20, it is reduced by around half (9.3) in the high-$T_c$ phase. This tendency is the same as in the BiS$_2$ system[19]. As mentioned above, the BiS$_2$ and the BiSe$_2$ systems showed very similar behavior under high pressure. It is suggested that these materials exhibit the same mechanism for the pressure induced phase transition. In the previous study of LaO$_{0.5}$F$_{0.5}$BiS$_2$, a pressure induced structural transition from tetragonal phase (*P4/nmm*) to monoclinic phase ($P2_1/m$) was suggested from XRD analysis under high pressure [16]. Such a kind of structural transition is one of the possible explanations for the pressure induced phase transition of $LaO_{0.5}F_{0.5}BiSe_2$, since it has the same structure as $LaO_{0.5}F_{0.5}BiS_2$ in low-$T_c$ phase. Currently, high-pressure XRD measurements for the $BiSe_2$ system have not yet been performed. Therefore, a detailed investigation for the structure of $BiS_2$ and $BiSe_2$ system may provide useful information for understanding the mechanism behind the high-$T_c$ phase.

Within recent days, a paper about $LaO_{0.5}F_{0.5}BiSe_2$ was submitted to arXiv [23]. it shows that the results of the resistivity measurements under high pressure are slightly different from our results. Although the pressure induced phase transition is also shown in that paper, in the low-$T_c$ phase, the $T_c$ gradually decreases with increasing applied pressure [23]. This different behavior in low-$T_c$ phase may be due to the difference of the fluorine concentration between our samples and the reported samples.

In summary, we revealed that the $T_c$ of $LaO_{0.5}F_{0.5}BiSe_2$ discontinuously increases. Although our sample does not show zero resistivity at ambient pressure, by applying pressure up to 2.44 GPa, maximum $T_c^{onset}$ and $T_c^{zero}$ were observed at 6.7 K and 5 K. The resistivity of both low- and high-$T_c$ phases shows metallic behavior. This is very different from what was seen in the $BiS_2$ system. The $\gamma$ extremely decreases from 20 to 9.3 after the phase transition, and the resistivity of the normal conducting state in high-$T_c$ phase gradually increases with increasing applied pressure which is in accordance with the $BiS_2$ system. In this current stage, the optimal $T_c$ of $LaO_{0.5}F_{0.5}BiSe_2$ was not observed even at the maximum pressure obtained during this study. Therefore applied further pressure has the potential to induce a much higher $T_c$ in this system.

This work was supported in part by the Japan Science and Technology Agency through Strategic International Collaborative Research Program (SICORP-EU-Japan) and Advanced Low Carbon Technology R&D Program (ALCA) of the Japan Science and Technology Agency.

FIG. 1

XRD patterns of $LaO_{0.5}F_{0.5}BiSe_2$. All diffraction peaks are assigned to the (00l) indices.

FIG. 2

(a): Resistivity versus temperature from 300 K to 2 K for $LaO_{0.5}F_{0.5}BiSe_2$ under varying pressure. (b): The expanded view around $T_c$.

FIG. 3

Resistivity versus temperature at 1.17 and 2.24 GPa under magnetic field with the different directions, which are parallel with *ab* plane (*H*//*ab*) and *c* axis (*H*//*c*).

FIG.4

Upper critical field versus temperature for the low-$T_c$ (black) and high-$T_c$ (red) phases. Solid circles denote the $H_{c2}^{//ab}$. Open circles denote $H_{c2}^{//c}$. The dashed lines denote the straight-line approximation.

TABLE I. Superconducting parameters for $La(O_{0.5}F_{0.5})BiSe_2$ under 1.17 (low-$T_c$ phase) and 2.44 GPa (high-$T_c$ phase).

| Pressure (GPa) | $dH_{c2}^{//ab}/dT$ (T/K) | $dH_{c2}^{//c}/dT$ (T/K) | $H_{c2}^{//ab}(0)$ (T) | $H_{c2}^{//c}(0)$ (T) | $\xi_{ab}(0)$ (nm) | $\xi_c(0)$ (nm) | $\gamma(0)$ |
|---|---|---|---|---|---|---|---|
| 1.17 | -1.94 | -0.01 | 5.5 | 0.6 | 23.4 | 2.56 | 20 |
| 2.44 | --3.39 | -0.59 | 20.6 | 3.4 | 9.83 | 1.63 | 9.3 |

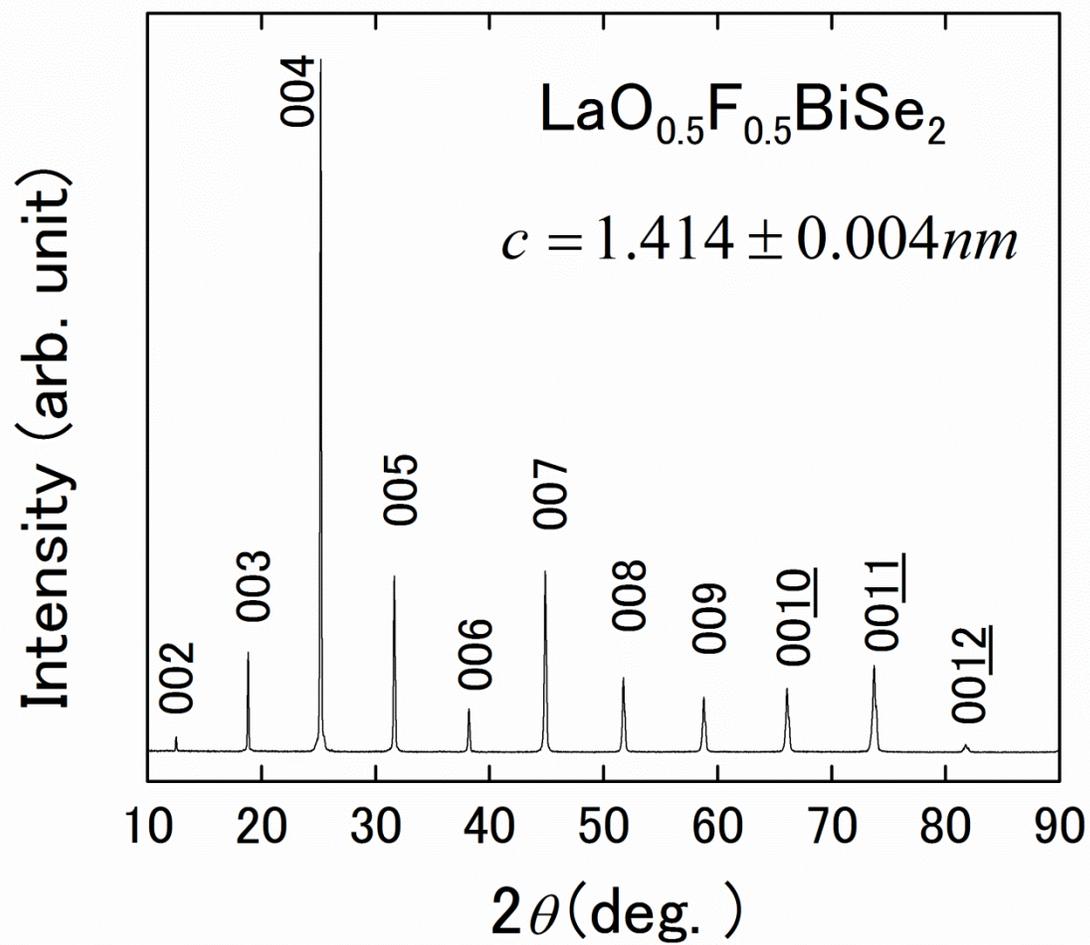

Fig. 1

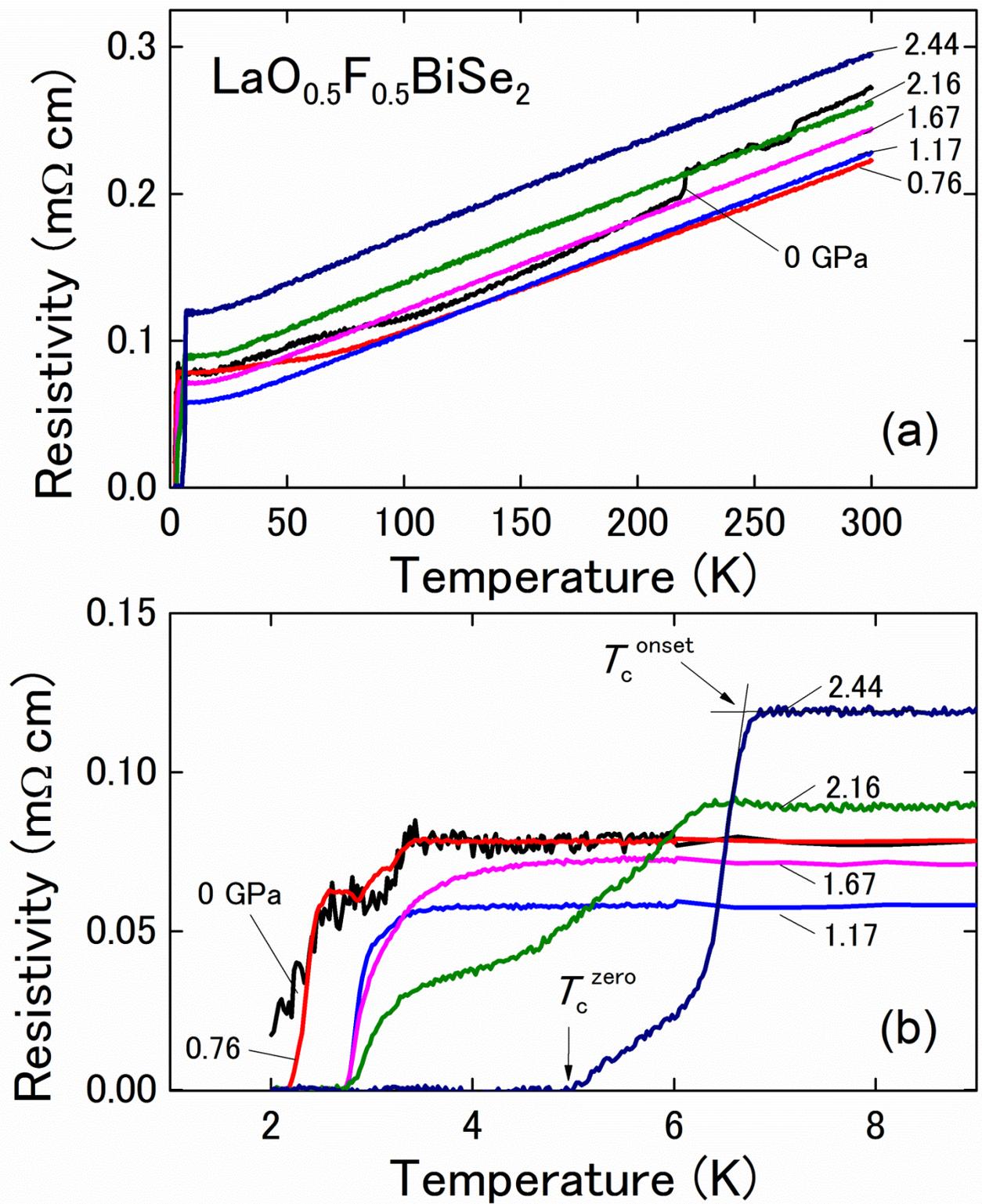

Fig. 2

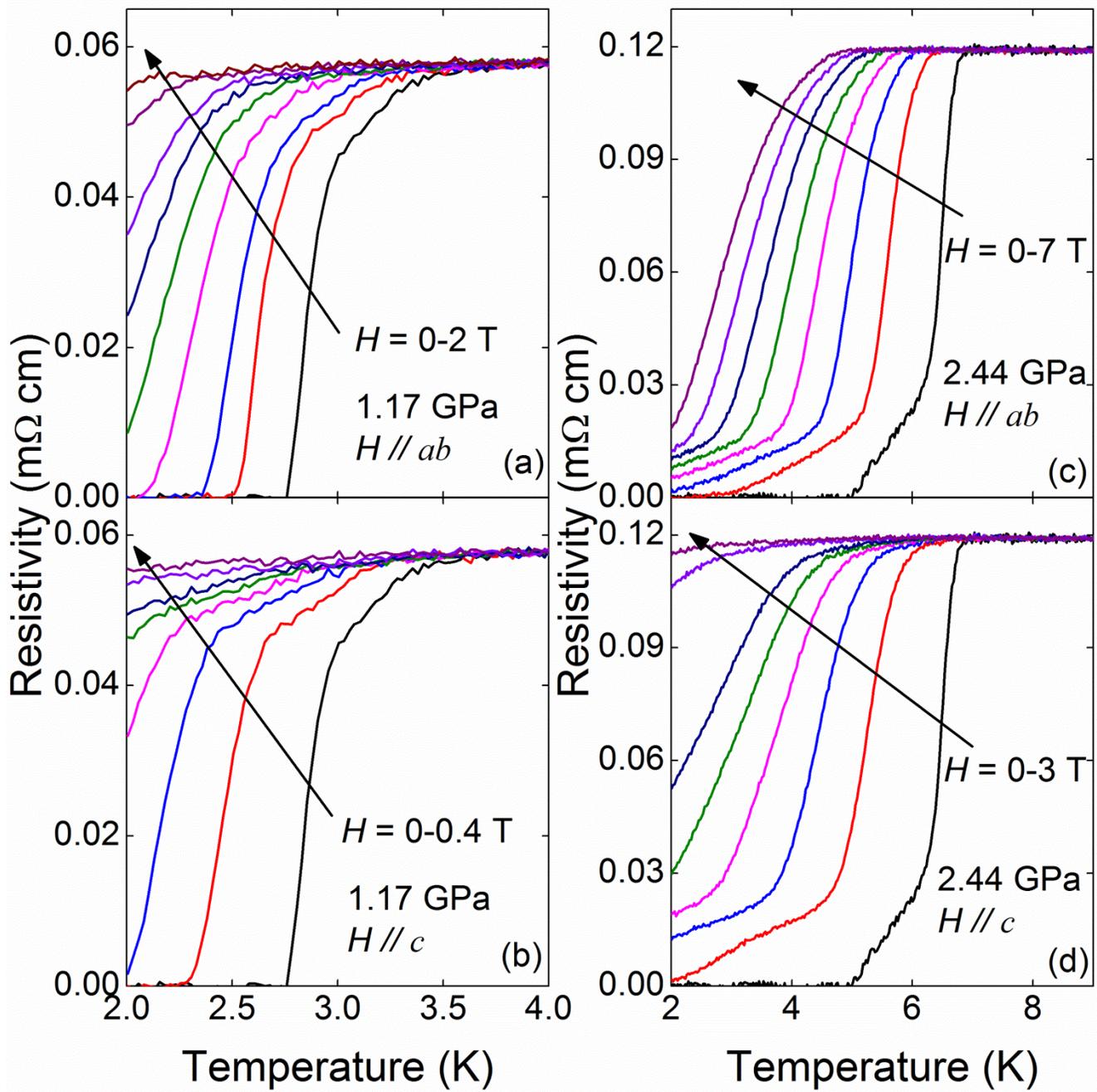

Fig. 3.

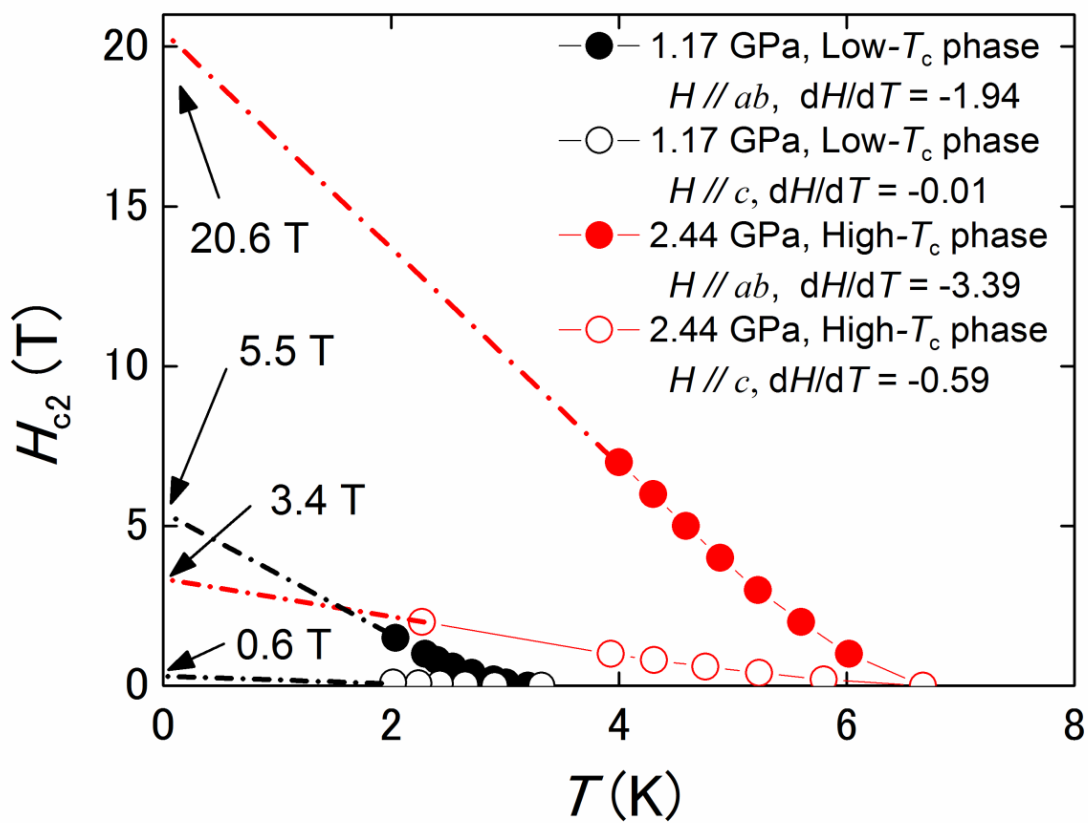

Fig. 4.